\documentclass[reprint,superscriptaddress,amsmath,amssymb,aps]{revtex4-2}

\usepackage{graphicx}
\usepackage{dcolumn}
\usepackage{bm}
\usepackage{ulem}
\usepackage{bm}
\usepackage{algorithm}
\usepackage[noend]{algpseudocode} 
\usepackage{amssymb}
\usepackage{soul}
\usepackage[mathlines]{lineno}
\usepackage{tabularx}
\usepackage{graphicx}
\usepackage{dcolumn}
\usepackage{bm}
\usepackage{amsmath}
\usepackage{xcolor} 
\usepackage[pagebackref]{hyperref} 
\hypersetup{
	colorlinks   = true, 
	urlcolor     = blue, 
	linkcolor    = blue, 
	citecolor   = blue 
}
\usepackage{graphicx}
\usepackage{physics}
\usepackage{mathtools}
\usepackage{dsfont}
\usepackage{natbib}

\begin{document}
\title{Propagation invariant one-dimensional flat-top vector beam with conserved topology}

\author{Cathy Maako}
\address{School of Physics, University of the Witwatersrand, Private Bag 3, Wits 2050, South Africa}

\author{Yuancong Cao}
\address{School of Physics, Xidian University, Xi'an 710071, China}

\author{Pedro Ornelas}
\email{1836488@students.wits.ac.za}
\address{School of Physics, University of the Witwatersrand, Private Bag 3, Wits 2050, South Africa}

\author{Mingjian Cheng}
\address{School of Physics, Xidian University, Xi'an 710071, China}

\author{Andrew Forbes}
\address{School of Physics, University of the Witwatersrand, Private Bag 3, Wits 2050, South Africa}
\author{Angela Dudley}
\email{Angela Dudley@wits.ac.za}
\address{School of Physics, University of the Witwatersrand, Private Bag 3, Wits 2050, South Africa}

\author{Isaac Nape}
\email{isaac.nape@wits.ac.za}
\address{School of Physics, University of the Witwatersrand, Private Bag 3, Wits 2050, South Africa}

\vspace{10pt}



\begin{abstract}
\noindent \textbf{We report the generation and control of a propagation invariant one-dimensional flat-top vector beam through the digitally controlled superposition of a first-order petal- and Gaussian- mode each marked with orthogonal polarizations. We present a detailed analysis of the spatial and polarization profiles and unveil their intricate topological features, showing that they have properties that are invariant to the propagation dynamics, in addition to their preserved intensity. The differential Gouy phase between the spatial components is shown to rotate the great circle traced by the polarisation distribution on the Poincaré sphere. Finally, we assess the beams for laser cutting and optical trapping, highlighting the relevance of topological structured light to these applications.}
\end{abstract} 
\vspace{2pc}
\maketitle
\section*{Introduction}
Structured light enables the tailoring of optical fields through control of their wavelength, spatial amplitude and phase, and polarisation \cite{forbes2021structured, rubinsztein2017roadmap}. In particular, simultaneous control of the polarisation and spatial components of light has enabled the realisation of optical fields with tailored polarisation textures carrying singularities \cite{dennis2009singular, angelsky2022optical}, knots \cite{larocque2018reconstructing, dennis2010isolated}, links \cite{zhou2026spatiotemporally}, Möbius strips \cite{bauer2015observation, forbes2021structured}  and skyrmionic topologies \cite{shen2022generation, shen2024optical}. This capability is supported by an expanding toolbox that includes tailored metasurfaces and engineered materials \cite{dorrah2022tunable, wang2021generation}, static liquid-crystal devices such as q-plates\cite{rubano2019q}, dynamic liquid-crystal devices such as spatial light modulators (SLMs) \cite{rosales2024structured, yang2023review}, and digital micromirror devices \cite{scholes2020structured, perez2022highly}.\\
Beyond creating complex polarisation textures, the joint control of polarisation and spatial mode shaping provides a means of engineering the intensity profile of light and its evolution during propagation. This principle has enabled the generation of vector flat-top beams whose transverse intensity shape is preserved during propagation, apart from an overall change in scale \cite{bhebhe2018classical, han2011flattop}. This is achieved by combining complementary spatial profiles in orthogonal polarisation components where their intensities can be added without interference to produce a desired beam shape. Such designs address the limitation of conventional flat-top shaping approaches that produce the desired profile only near a prescribed plane, with demonstrated applications in optical trapping and tweezing \cite{kritzinger2022optical}. However, these existing vector flat-top designs focused on cylindrically symmetric intensity profiles. Extending this approach beyond cylindrical symmetry, while preserving the transverse intensity shape during propagation and enabling flexible control of its geometry and orientation, remains unexplored. 

Here, we address this limitation by presenting a one-dimensional flat-top vector beam that preserves both its intensity profile and polarisation topology under free-space propagation. Unlike conventional flat-top approaches, which rely on fixed optical elements and produce profiles valid only near a prescribed plane, our approach exploits the digital control afforded by a spatial light modulator to generate and reconfigure the beam in real time. The resulting field is nonseparable in its spatial and polarisation degrees of freedom, and we show that its polarisation structure traces a stable great circle on the Poincaré sphere during propagation - a topological feature characterised by a conserved integer winding number. We demonstrate the practical utility of this control through applications in laser cutting and optical trapping.


\begin{figure*}[hbt!]
    \centering
    \includegraphics[width = \linewidth]{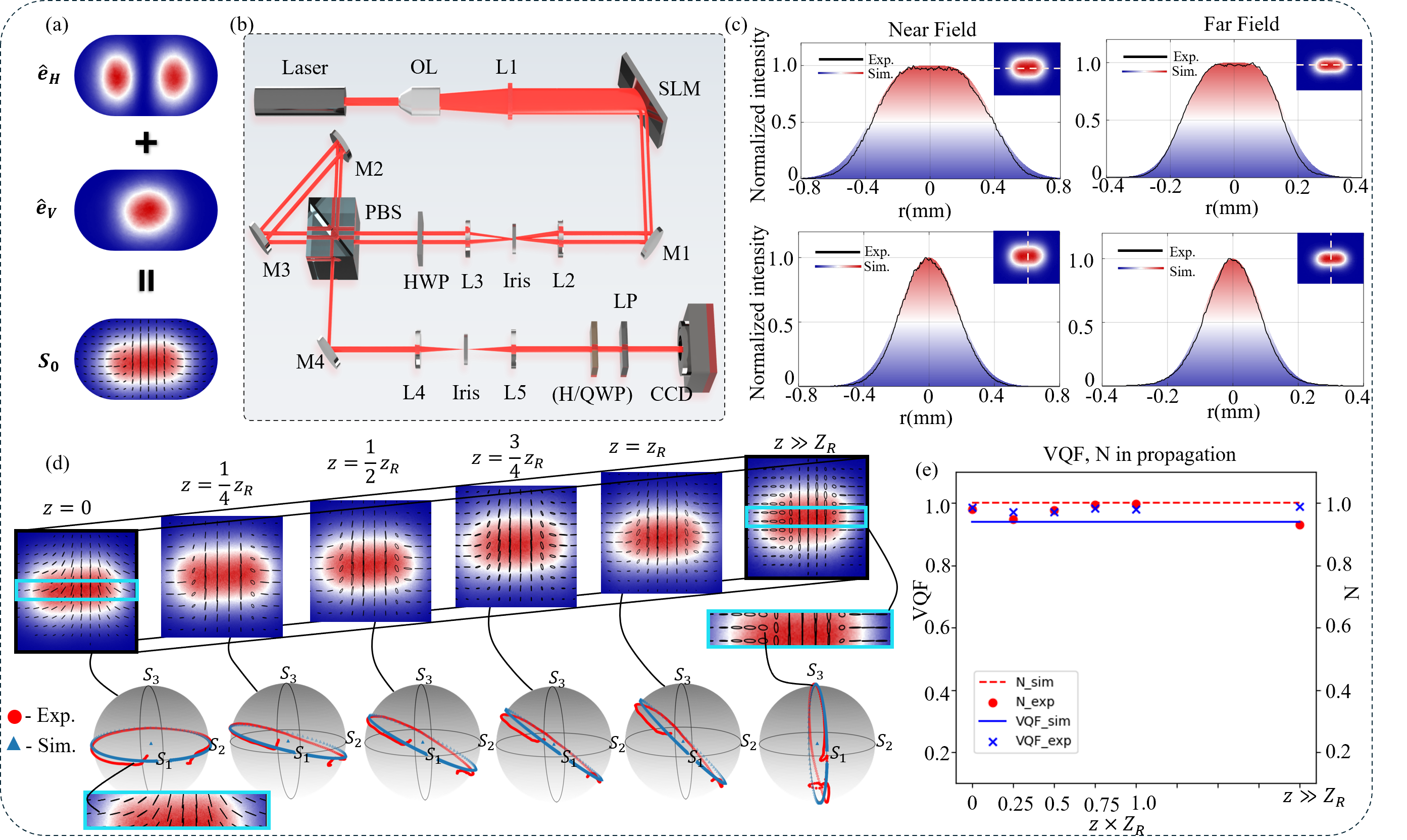}
    \caption{(a) Generation principle of a one-dimensional vector flat-top beam.  A horizontally polarised ($\boldsymbol{\hat{e}_H}$) petal mode is coherently superposed with a vertically ($\boldsymbol{\hat{e}_V}$) polarised Gaussian mode to produce a vector flat-top beam with a spatially inhomogeneous polarisation distribution. (b) Experimental setup for generating and characterising the vector beam. The generation scheme consists of lenses (L) for beam expansion onto a Spatial light modulator (SLM) for independent beam generation and a Sagnac interferometer consisting of 2 mirrors (M) and a polarizing beam splitter (PBS) to superimpose the two orthogonally polarized beams. A set of polarisation optics, including half-wave plates (HWP), quarter-wave plate (QWP), and a polariser, enable Stokes polarimetric analysis at different propagation planes. 
    (c) Experimental (black line plot) and numerically simulated (surface plots) near- and far- cross-sectional profiles of field intensity distributions (shown as insets), showing a flat-top distribution along the horizontal direction and a Gaussian distribution along the vertical direction. (d) Spatially-varying state of polarization (SOP) represented by local polarisation ellipses and corresponding polarisation state population plotted on the Poincaré sphere, reconstructed from the Stokes parameters, 
    at propagation distances expressed relative to the Rayleigh range, $z_R$. (e) Measured topological invariant $N$ and nonseparability, quantified by the vector quality factor (VQF), at each observation plane.}
    \label{Figure 1}
\end{figure*}
\section*{Theory}
We generated a one-dimensional flattop by superimposing natural modes of free space, i.e., Laguerre–Gaussian beams that are orthogonally polarized, as shown in Fig.~\ref{Figure 1} (a) at the z=0 plane, producing the field
\begin{equation}
    \begin{split}
        U(\bar{r})&=\frac{\alpha}{\sqrt{2}}[\text{LG}^0_{1} \left(\bar{r}\right) + \text{LG}^0_{-1} \left(\bar{r}\right)]\boldsymbol{\hat{e}_H} + \sqrt{1-\alpha^2}\text{LG}^0_0(\bar{r}) \boldsymbol{\hat{e}_V}, \\
        & = \text{LG}^0_0(\bar{r}) \left( 2\alpha \ \frac{r \cos(\phi)}{w_0} \boldsymbol{\hat{e}_H} + \sqrt{1-\alpha^2}\boldsymbol{\hat{e}_V} \right)
    \end{split}
    \label{Field eq}
\end{equation}
Here $\bar{r}$ is the transverse coordinate vector, expressed as $\bar{r} = (r, \phi)$ in polar coordinates or $\bar{r} = (x, y)$ in Cartesian coordinates. The unit vectors $\boldsymbol{\hat{e}_H}$ and $\boldsymbol{\hat{e}_V}$ represent the horizontal and vertical polarisation states, which carry the superposition state $\frac{\mathrm{LG}^0_1 + \mathrm{LG}^0_{-1}}{\sqrt{2}} \propto \exp(-r^2/w_0^2)\frac{r}{w_0}\cos(\phi)$ and the Gaussian mode $\mathrm{LG}^0_0 \propto \exp(-r^2/w_0^2)$, respectively, where $w_0$ is the second-moment radius of the Gaussian envelope. The intensity profiles of these components and the resulting polarisation profile are shown in Fig.~\ref{Figure 1} (a). The horizontally polarised superposition resembles a petal mode, with two intensity peaks at $x=\pm \frac{\sqrt{3\alpha^2-1}w_0}{2\alpha}$ along the $x$-axis. Here $\alpha$ is the modal weighting parameter, where the flat-top condition is satisfied when $\alpha = \frac{1}{\sqrt{3}}$, which equates the peak intensity of the petal mode to that of the Gaussian mode along the flat-top axis (see Supplement for derivation).
However, because the two polarisation components have orthogonal spatial modes, the resulting field contains an inhomogeneous (spatially varying) polarisation profile.
\section*{Experimental realisation and results}
To generate this field, we used the experimental setup shown in Fig.~\ref{Figure 1} (b). A coherent laser beam was expanded onto a phase-only spatial light modulator (SLM) and structured using complex amplitude modulation to produce the desired beam profiles: the $\mathrm{LG}^0_{\pm1}(\bar{r},z=0)$ superposition on one half of the SLM and the $\mathrm{LG}^0_0(\bar{r},z=0)$ mode on the other. The fields were separated from the zero diffraction order by spatial filtering in the far-field, and their polarisation was rotated to diagonal; subsequently, the desired vector beam was generated by superimposing both beams using a Sagnac interferometer (see Ref \cite{wang2016sagnac}). Due to renormalisation of the SLM holograms and losses in the interferometer, the relative mode powers were readjusted by changing the grating depth to satisfy the peak-intensity matching condition, i.e. $\alpha = \frac{1}{\sqrt{3}}$. Furthermore, imaging lenses mapped the fields from the SLM plane ($z=0$) onto the CCD camera, with half-wave plates (HWPs), quarter-wave plates (QWPs) and a polariser placed before the camera for polarisation and intensity analysis.
To confirm the one-dimensional flat-top profile, we analysed the total beam intensity cross-sections along the Cartesian $x$ and $y$ directions in both the near field ($z=0$) and the far field (at the focus of a lens). The measured profiles (lines) and scaled theoretical/simulated profiles (shaded regions) are shown in Fig.~\ref{Figure 1} (c), revealing a flat-top profile along $x$ and a Gaussian profile along $y$. This contrasts with the two-dimensional flat-top field generated by Bhebhe \textit{et al}. \cite{bhebhe2018classical}, which used a doughnut mode ($\mathrm{LG}^0_1$) in the horizontal component instead of a petal mode.\\
\noindent Next we study the propagation-dependent dynamics and evolution of the polarisation structure. To do this, we measured the field at several propagation distances relative to the Rayleigh range of the Gaussian envelope, $z = n z_R$, where $n=0, 1/4, 1/2, 1$, followed by a final measurement in the far field. At each plane, Stokes polarimetry (see Supp.) was used to measure the local Stokes vector, $\bar{S}(\bar{r}, z)$, and total intensity, 
following the methods of Singh \textit{et al}. \cite{singh2020digital}. The intensity profiles, overlaid with the polarisation ellipses representing the local state of polarization (SOP), are shown in the top panel of Fig.~\ref{Figure 1} (d), revealing that the polarisation structure evolves upon propagation despite the one-dimensional flat-top profile being maintained. In the proceeding sections, we study our field's polarization behaviour in more detail.
\subsection*{Nonseparability analysis}
To quantify the vectorial characteristics of the one-dimensional flat-top beam, we measured its vector quality factor (VQF), a property that ranges from zero for a scalar field to unity for a maximally nonseparable spatial–polarisation field. Using 
the measured Stokes parameters obtained previously, we extracted the spatially integrated Stokes parameters, $\sigma_i=\int S_i,dA$, giving 
\begin{equation}
\mathrm{VQF}=\sqrt{1-\frac{\sigma_1^2+\sigma_2^2+\sigma_3^2}{\sigma_0^2}}.
\end{equation}
The theoretical value for the chosen mode weighting ($\alpha =\frac{1}{\sqrt{3}}$) is $\mathrm{VQF^{TH}}=0.94$, compared with a measured value of $\mathrm{VQF^{EXP}}= 0.989 \pm 0.010$  at $z=0$. Fig.~\ref{Figure 1} (e) presents the VQF measured at each plane of propagation,  
with the measured VQF remaining approximately constant within experimental uncertainty ($\Delta \mathrm{VQF} = 0.016$), 
showing that the nonseparability of the field remains preserved on propagation. It is important to note that since the VQF depends on the modal weighting $\alpha$, which determines the flatness of the beam profile, we can use the VQF as a quantitative measure of how well the flattop profile is retained under different transformations. We showed here that the flattop beam maintains a constant VQF in propagation; therefore, it follows that the flattop profile is invariant through propagation.

\subsection*{Polarisation evolution }
The preservation of the VQF of the field follows from the fact that it undergoes a unitary evolution in propagation. To understand this behaviour, we generalise ~\eqref{Field eq} to include the propagation dependence, $\mathrm{LG}^p_{l}(\bar{r}) \rightarrow \mathrm{LG}^p_{l}(\bar{r},z)$ for $l \in {0,\pm1}$. The field can therefore be written as (neglecting global phases)
\begin{equation}
    \begin{split}
        U(\bar{r}, z) = \text{LG}^0_0(\bar{r}, z) \left( 2\alpha \  \frac{r \cos(\phi)}{w(z)}  e^{i(\psi(z))} \boldsymbol{\hat{e}_H} + \sqrt{1-\alpha^2}\boldsymbol{\hat{e}_V} \right). 
    \end{split}
    \label{Field eq2}
\end{equation}
Both the petal (horizontal polarisation) and Gaussian (vertical polarisation) modes maintain their intensity profiles during propagation, with rescaled sizes, explaining the preserved intensity shape, but a relative phase $\psi(z) = \operatorname{arctan}(\tfrac{z}{z_R})$ between the horizontal and vertical polarisation components emerges, which explains the changing polarization states. This is due to the differential Gouy phase generated by the spatial components: the Gaussian mode acquires a Gouy phase of $\psi(z)$, while the petal mode acquires $2\psi(z)$. The latter inherits its Gouy phase from the $\mathrm{LG}_{\pm1}$ modes, which acquire identical Gouy phases, giving a common phase to the horizontal polarisation component. Thus, as shown in Fig.~\ref{Figure 1} (d), the relative phase between the two polarisation components drives the evolution of the polarisation structure while preserving the intensity profile.
\begin{figure*}[hbt!]
    \centering
    \includegraphics[width =\linewidth]{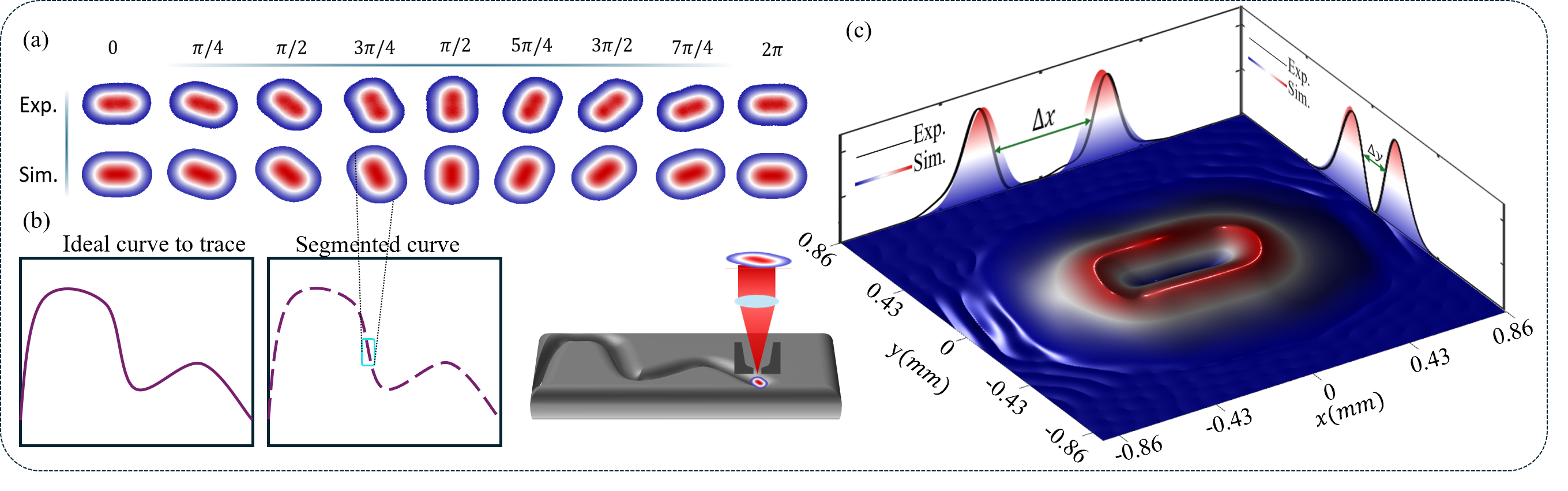}
    \caption{(a) Digital tuning of the flat-top orientation through intramodal phase, $\theta$, control 
    with experimental and simulated intensity profiles shown for selected $\theta$ values. (b) Application of digital orientation control for precise laser cutting along predefined segmented curves. (c) Trapping-force distribution inferred from the measured intensity of the 1D flat-top beam, showing a narrower trapping region along the Gaussian axis and a broader region along the flat-top axis. Insets show cross-sectional trapping-force profiles and the force vector field.} 
    \label{Figure 2}
\end{figure*}

 Next, we analysed the nature of the populated states during the evolution by mapping the measured fields onto the Poincaré sphere at each measurement plane ($z$) using the normalised Stokes parameters, i.e. $\hat{S}(\bar{r})  = \bar{S}(\bar{r}) / |\bar{S}(\bar{r})|$, s.t $\hat{S}(\bar{r}) \cdot \hat{S}(\bar{r}) =1$ at every point $\bar{r}$ where the fields are defined. The experimental and theoretical/simulated results are shown in the bottom panel of Fig.~\ref{Figure 1} (d). Given the cross-sectional symmetry of the polarisation profile about the $y$-axis, it was sufficient to extract the Stokes parameters from a neighborhood of points around the $y=0$ cross-section, across all $x$ positions where the intensity is defined. Both experiment and simulation show that, at $z=0$, the polarisation states form a closed loop along the equator of the Poincaré sphere. During propagation, this loop rotates about the $S_1$ axis while remaining closed, signaling a smooth deformation that does not change the intensity but also preserves the mapping onto a closed loop. To examine the topological nature of this mapping, we calculated the theoretical normalised Stokes parameters for the field in ~\eqref{Field eq2}, yielding the vector
\begin{equation}
 \hat{S}(\bar{r}, z) \propto 
 \begin{pmatrix}
      -\cos(f(x,z)) \\
      \sin(f(x,z)) \cos(2\psi(z))    \\
      \sin(f(x,z)) \sin(2\psi(z))    \\
 \end{pmatrix}
\end{equation}
where the function $f(x,z) = \tan^{-1}\left(\tfrac{4\mu x}{w(z)}\right)$ with $\mu = \sqrt{\tfrac{\alpha^2}{1-\alpha^2}}$ maps $-\infty<x<\infty$ to the domain $-\pi<f(x,z)< \pi$. The polarisation evolution corresponds to a rotation of the populated great circle on the Poincaré sphere through an angle twice the relative phase (differential Gouy phase) as the beam propagates. Accordingly, the normalised Stokes parameters thus define a mapping from the spatial coordinate, through $f(x,z)$, onto this great circle. Since the field satisfies $\hat{S}(x\rightarrow \mp \infty,z) = (1, 0, 0)$, the two spatial limits can be identified, satisfying the 
one-point compactification condition for the real line, $\mathbb{R}^1 \cup \{\infty\} \cong S^1$. The Gouy phase evolution continuously deforms the polarisation texture while keeping it constrained to a great circle. Within this constraint, the resulting $S^1\rightarrow S^1$ mapping admits an integer winding number, classified by the first homotopy group, $\pi_1(S^1)=\mathbb{Z}$, and is measured from (with more details given in the supplementary material) 
\begin{equation}
    N(z) =\frac{1}{2\pi} \oint_{S^1} \nabla \Phi_{xy}\cdot d\bar{\ell} ,
\end{equation}
where $d \bar{l}$ defines a segment on the great circle, tangent to each point, and $\Phi_{xy}$ is the Stokes phase defined as $\tan^{-1}\left(\tfrac{S_y}{S_x}\right)$, where $S_x$ and $S_y$ are chosen such that $S_z$ is defined orthogonal to the great circle. The measured and simulated results for this are shown in Fig.~\ref{Figure 1} (e), presenting an average winding number of $N(z)= 0.968$ for the measured planes with a standard deviation of $0.0292$, therefore confirming the topological stability with propagation. This winding number describes how many times the polarisation states wrap around the Poincar\'e sphere on the constrained great circle while the field evolves. Additionally, the sign of the winding number can be changed by performing a swapping operation of the polarization components in the field function (i.e. $\hat{e}_{H(V)} \rightarrow \hat{e}_{V(H)}$).  The sub-unity value of $N = 0.968$ is consistent with the expected numerical uncertainty arising from the discrete integration of the winding number along a finite path length, and from residual experimental noise in the measured Stokes parameters. It does not indicate a physical deviation from integer topology. The agreement between the measured and simulated winding numbers, both remaining constant across all observation planes, confirms the topological stability of the field. The derived 1D topology observed here arises as a result of the annihilation of optical Skyrmions with opposite vorticity \cite{teng2025optical}, which can be identified by partitioning Eq.~\ref{Field eq} into two distinct vector modes, each carrying a vertically polarized Gaussian component and distinct horizontally polarized LG components.

\subsection*{Analysis for laser cutting and optical trapping}
Next, we demonstrate the utility of our digital encoding approach of the one-dimensional flat-top for laser cutting and optical trapping. 
 In laser cutting applications, dynamic control of the beam can be advantageous in free-form laser ablation. We demonstrate how our digital approach enables this control through a single beam parameter, without mechanically or physically rotating any optics or a sample. The results for this are shown in Fig.~\ref{Figure 2} (a), where the flat-top is physically rotated in the x-y plane by controlling the intermodal phase, $\theta$, between the horizontally polarized LG modes, i.e. $\tfrac{1}{\sqrt{2}}[\text{LG}_{1}^0 \left(\bar{r}\right) + e^{i\theta}\text{LG}_{-1}^0 \left(\bar{r}\right)]$. Here, the inter-modal phase is varied from $0$ to $2\pi$, where the axis in which the flattop profile exists rotates with changing $\theta$. Given an SLM with bit depth $D = 256$, the achievable resolution in rotation can reach up to $\Delta \theta = 1/D \times 2\pi = 0.0245$ and can be increased using SLMs with higher bit depths. Moreover, the orientation of the field is independent of the relative phase between the polarisation components (as was seen in the propagation analysis), making it insensitive to fluctuations in the interferometer. Since control over the intermodal phase is achieved through digital encoding on the SLM, our method offers an easily accessible degree of freedom for flattop beam shaping to achieve more diverse optical traps and enhanced control of laser cutting configurations. The ability to easily tune the phase and rotate the beam can be useful in creating a laser cutting system that can be used on both thin materials for fine detailed cuts and on thicker materials. Moreover, this rapid tunability allows one to trace out more complex and detailed curves to cut, as shown in Fig.~\ref{Figure 2} (b). 
Flat-top beams and Gaussian beams have both been well researched for applications in optical trapping and laser cutting, both offering unique advantages and disadvantages depending on the application \cite{thomas2026laser,kritzinger2022optical, nacius2024polarization, hao2008smallest}.
In laser cutting applications, the beam profile influences the cut quality, surface and edge finish, power distribution, and more \cite{alsaadawy2024comprehensive}. Gaussian beams have their intensity concentrated around a single peak; therefore, one can achieve a tighter focus and higher peak intensity, which is useful for fine, detailed cutting on thin objects. However, the Gaussian profile decays rapidly away from the center; this will affect cut quality on thicker objects and will create more slanted kerfs. Flattop beams have uniform intensity across a region and have steeper edges; therefore, the power distribution is more consistent through the cut and leaves straighter kerfs and higher-quality cuts on thicker objects. Because our 1D flattop beam has both Gaussian and flattop profiles in different axes, one can achieve laser cutting that contains the advantages of both beam profiles. 
For uses in optical trapping, we analyse the force field of an optical trap from the intensity of the beam,$F_{t} = c\nabla I $, where the constant $c$ depends on the properties of the trapped particle and the medium of the trap \cite{neuman2004optical, yang2021optical}. Fig.~\ref{Figure 2} (c) shows the experimental and simulated plots of the magnitude of the force field of our 1D flattop, thereby highlighting its trapping landscape. In an optical trap setup, a particle would be pushed to the central region of low strength. More interestingly, we observe that the field has a much wider trap region along the axis with a flat-top profile and a sharper trap region along the axis with a Gaussian profile, $\Delta x > \Delta y$. The experimental cross sections (lines) and the theoretical cross-sections (shaded region) are shown in Fig.~\ref{Figure 2} (c). This may be useful for trapping particles more tightly in one axis and allowing them to move more freely in another axis. Moreover, because the orientation of the flat-top can be digitally controlled, one can manipulate the trajectory of the trapped particles in any direction on the x-y plane.\\
\section*{Discussion and conclusion}
Here we have engineered a one-dimensional vector flat-top beam by superimposing propagation-invariant spatial patterns marked with orthogonal polarisations. We showed that the resulting field preserves its intensity and orientation, remaining a flat-top along a single axis, during propagation. Furthermore, we quantified the vectorial nature of the beam using its VQF and identified that its polarization distribution forms a great circle on the Poincaré sphere. Complementary to the 1D flat-top nature of the field, we identified that the polarization great circle can be identified along the flat-top axis, thereby forming the mapping $R^1 \to S^1$ that can be characterized by a well-defined integer topological winding number $N=1$. Furthermore, we found that the polarization great circle would rotate in propagation due to propagation induced Gouy-phase while the VQF and $N$ would remain unchanged. Finally, by demonstrating control over the 1D flat-top axis, we proposed the usage of such beams for laser cutting over predefined paths and furthermore, by quantifying the trapping force of the field, we identified its inherent asymmetric trapping landscape, which may allow for more diverse trapping capabilities. The approach demonstrated here is not limited to the specific mode superposition employed, and may be extended to other mode families or higher-dimensional flat-top profiles, opening a route toward reconfigurable vectorial beam shaping for a broad range of structured-light applications.\\

\subsection*{Funding} C.M. acknowledges support from the South African Quantum Technology Initiative (SA QuTI). P.O. acknowledges support from the DSTI-CSIR Inter-programme Bursary Scheme. I.N. and A.D. acknowledge support from the DSI-CSIR Rental Pool Programme. A.F. acknowledges support from the Oppenheimer Memorial Trust (OMT), the South African Quantum Technology Initiative (SA QuTI) and the DSI-CSIR Rental Pool Programme. Y.C. and M.C. acknowledge support from the National Natural Science Foundation of China (Grant No. 62575227, 62231021, U254120011) and Shaanxi Project for Basic Scientific Research in Mathematics and Physics (Grant No. 25JSQ005).

\subsection*{Acknowledgments} The authors would like to acknowledge Lehloa Mohapi for his initial work on the project.\\
\subsection*{Disclosures} The authors have no conflicts of interest to declare\\
\subsection*{Data availability} The underlying data for the results presented in this paper are not publicly available. The data may be obtained from the authors upon reasonable request.\\


\clearpage
\appendix

\setcounter{section}{0}
\setcounter{figure}{0}
\setcounter{table}{0}
\setcounter{equation}{0}
\setcounter{footnote}{0}
\renewcommand{\thesection}{S\arabic{section}}
\renewcommand{\thefigure}{S\arabic{figure}}
\renewcommand{\thetable}{S\arabic{table}}
\renewcommand{\theequation}{S\arabic{equation}}
\begin{widetext}
    
\section*{Supplementary: Derivation of coefficients}
The flat-top field at plane z = 0 is given by
\begin{equation}
    \begin{split}
        U(\bar{r})&=\frac{\alpha}{\sqrt{2}}[\text{LG}^0_{1} \left(\bar{r}\right) + \text{LG}^0_{-1} \left(\bar{r}\right)]\boldsymbol{\hat{e}_H} + \sqrt{1-\alpha^2}\text{LG}^0_0(\bar{r}) \boldsymbol{\hat{e}_V}, \\
        & = \text{LG}^0_0(\bar{r}) \left( 2\alpha \ \frac{r \cos(\phi)}{w_0} \boldsymbol{\hat{e}_H} + \sqrt{1-\alpha^2}\boldsymbol{\hat{e}_V} \right)
    \end{split}
    \label{Field eq init}
\end{equation}
where the unit vectors $\boldsymbol{\hat{e}_H}$  and $\boldsymbol{\hat{e}_V}$  represent the horizontal and vertical polarisation states, respectively; $\alpha$ determines the weighting of the two beams on the total field; and $\text{LG}_l(\bar{r})$ is the Laguerre-Gaussian mode equation given as 
\begin{equation}
    \begin{split}
        LG_l^p &= \sqrt{\dfrac{2p!}{\pi(|l|+p)!}}\dfrac{1}{w_0}\left[\dfrac{\sqrt{2}r}{w_0}\right]^{|l|}L_p^{|l|}\left[\dfrac{2r^2}{w_0^2}\right]\exp{\left(\tfrac{-r^2}{w_0^2}\right)}\\
        &\times\exp{\left(-il\varphi\right)}\exp{\left(-ik\dfrac{r^2}{2R(z)}\right)}
         \exp{\left(i\psi(z)\right)}.
    \end{split}
    \label{LG eq}
\end{equation}
Here, the $l$ and $p$ represent the azimuthal and radial indices, $w_0$ is the encoded beam waist, $L_p^{|l|}$ is the generalised Laguerre polynomials, and $\psi(z)= (|l| + 2p +1)\operatorname{arctan}(\tfrac{z}{z_R})$ is the Gouy phase acquired by the beam through propagation.\\ To derive the appropriate coefficient $\alpha$ for a flat-top beam, we consider the simplest case, $y = 0$, such that the field reduces to 

\begin{equation}
    \begin{split}
        U(\bar{r})
        & = \sqrt{\dfrac{2}{\pi w_0^2}} \operatorname{exp}\left(-\dfrac{x^2}{w_0^2}\right)\left( 2\alpha \ \frac{x}{w_0} \boldsymbol{\hat{e}_H} + \sqrt{1-\alpha^2}\boldsymbol{\hat{e}_V} \right)
    \end{split}
    \label{Field eq}
\end{equation}
and the intensity of the whole field becomes 
\begin{equation}
    \begin{split}
        I_{t} = \dfrac{2}{\pi w_0^2} \operatorname{exp}\left(-\dfrac{2x^2}{w_0^2}\right)\left( 4\alpha^2 \ \frac{x^2}{w_0^2} + 1-\alpha^2 \right)
    \end{split}
\end{equation}

From this, we can calculate the positions at which the intensity is maximum using the first derivative of the intensity
\begin{equation}
    \begin{split}
         \dfrac{dI_t}{dx} &=-\frac{8x}{\pi w_0^3}\operatorname{exp}\left(-\frac{2x^2}{w_0^2}\right)\left[\frac{4\alpha^2x^2}{w_0^2} -3\alpha^2 +1\right]\\
         \dfrac{dI_t}{dx} &=0\\
    \end{split}
\end{equation}
Solving for x, we get 3 roots
\begin{equation}
    \begin{split}
        x_1 &= 0\\
        x_2 &= \frac{\sqrt{3\alpha^2-1}w_0}{2\alpha}\\
        x_3 &= -\frac{\sqrt{3\alpha^2-1}w_0}{2\alpha}
    \end{split}
\end{equation}

We can use 2 of the roots to find the coefficient $\alpha$. For a flattop, we ideally want the peak intensity of the Gaussian, which is at $x_1 = 0$, to be equal to the peak intensity of the superposition mode, which will be at $x_2$ and $x_2$; therefore, to solve for $ \ alpha$, we can take
\begin{equation}
    \begin{split}
     \dfrac{4\alpha^2}{\pi w_0^2} \operatorname{exp}\left(-\frac{(3\alpha^2-1)}{2\alpha^2}\right) - \dfrac{2(1-\alpha^2)}{\pi w_0^2} &= 0\\
     \alpha &= \frac{1}{\sqrt{3}}
    \end{split}
\end{equation}

\section*{Supplementary: Evolution of polarization and topology in propagation}

\noindent As given by Eq. (3) in the main text, the field under propagation can be written as (neglecting global phases)
\begin{equation}
    \begin{split}
        U(\bar{r}, z) = \text{LG}_0(\bar{r}, z) \left( 2\alpha \  \frac{r \cos(\phi)}{w(z)}  e^{i(\psi(z))} \boldsymbol{\hat{e}_H} + \sqrt{1-\alpha^2}\boldsymbol{\hat{e}_V} \right). 
    \end{split}
    \label{Field eq2}
\end{equation}

\noindent Locally normalizing the field such that 
$|U(\bar{r}, z)|^2 = 1,\; \forall (\bar{r},z)\in R^3$ and rewriting in cartesian coordinates the above expression can be further simplified to 

\begin{equation}
    \begin{split}
        u(x', z) =  \frac{U(x', z)}{|U(x', z)|^2} = \frac{1}{\sqrt{1+x'^2}}\left( x' e^{i(\psi(z))} \boldsymbol{\hat{e}_H} + \boldsymbol{\hat{e}_V} \right), 
    \end{split}
    \label{Field eq3}
\end{equation}

where $x' = \frac{2}{w(z)}\sqrt{\frac{\alpha^2}{1-\alpha^2}}x$ is the normalized $x$-coordinate scaled under propagation with the assumption that $\alpha \in (0,1)$. In this form, it is clear that the field has no dependency on the $y$-coordinate, with relative amplitude changes occurring between the two polarized components as we sweep across the $x$-coordinate. At $z=0$, sweeping across $x$ results in a rotation of the polarization state from vertical at $x'=0$ to diagonal at $x'=1$ and anti-diagonal at $x'=-1$ and finally to horizontal as $x'\to\pm\infty$. In propagation the field accumulates relative phase between the two orthogonally polarized components. The relative phase increases monotonically as $\psi(z) = \tan(z/z_R)$ with $z\in[0,\infty)$. As the field is allowed to propagate to the far-field ($z\to\infty$ or $z>\!\!>z_R$) we have that $\psi(z)\to \pi/2$, thereby altering the polarization behaviour across $x'$. Whilst the polarization at the points $x'=0$ and $x\to\pm\infty$ remain fixed we find that at $x'=\pm1$ we now find left and right circular polarization, respectively. Therefore, whilst the propagation leaves the boundaries of the field unchanged in both polarization and intensity (as discussed in main text), the evolution of the polarization states change in propagation, tracing a great circle over the Poincar\'e sphere that rotates upon propagation, as further evidenced by Fig. 1(d) in the main text.\\

\noindent To analyze the topological structure of the field, we can compute the locally normalized Stokes parameters which can be extracted directly from the field using the Hopf map, 
\begin{equation}
    \vec{S} = \left[|U_H|^2 - |U_V|^2, 2\text{Re}\left( U_H U_V^*\right), 2\text{Im}\left( U_H U_V^*\right)\right]^T
\end{equation}, thereby yielding

\begin{equation}
        \vec{S}(x',z) = \frac{1}{1+x'^2}\begin{bmatrix}
            x'^2 - 1 \\
            x' \cos(\psi(z))\\
            x' \sin(\psi(z)) \\
        \end{bmatrix}.
    \label{StokesNorm1}
\end{equation}

If we let $x' = \tan(f(x'))$ then we can simplify the above expressions using the identities $\cos(2\theta) = \frac{1-\tan^2\theta}{1+\tan^2\theta}$ and $\sin(2\theta) = \frac{2\tan\theta}{1+\tan^2\theta}$ to

\begin{equation}
        \vec{S}(x',z) = \begin{bmatrix}
            -\cos(2f(x')) \\
            \sin(2f(x')) \cos(\psi(z))\\
            \sin(2f(x')) \sin(\psi(z)) \\
        \end{bmatrix}.
    \label{StokesNorm2}
\end{equation}

Here we are interested in studying the topological structure of our field and its evolution as the field propagates. It is simplest therefore to study the field at $z=0$ and then consider the transformation enacted upon propagation. Therefore at $z=0$, the Stokes parameters are given simply by
\begin{equation}
        \vec{S}(x') = \begin{bmatrix}
            -\cos(2f(x')) \\
            \sin(2f(x'))\\
            0\\
        \end{bmatrix}.
    \label{StokesNorm3}
\end{equation}

At $x' = 0$ the above expression evaluates to $\vec{S}(x') = [-1, 0, 0 ]^T$ and as $x' \to \pm \infty$ we have $\vec{S}(x') = [1, 0, 0 ]^T$. Therefore,  $\vec{S}(x')$ is a map between $R^1 \to S^1$ where $R^1\cup\{\infty\}$ can be identified with a circle, $S^1$, under one point compactification of the points at infinity, $x\to\pm\infty$. The topological winding number, $N_0$, of this map can be computed using 
\begin{align}
    N^{12}_0 = \frac{1}{2\pi}\oint_{S^1} \nabla \Phi_{12}\cdot d\bar{\ell},
\end{align}

where $\Phi_{12} = \tan^{-1}\left(\frac{S_1}{S_2}\right)$. Using $\nabla \Phi_{12} = \frac{S_2\partial_x S_1 - S_1\partial_x S_2}{S_1^2 + S_2^2}$, $\partial_{x'}S_2 = 2\cos(2f(x'))\cdot \frac{\partial f(x')}{\partial x'}$ and $\partial_{x'}S_1 = 2\sin(2f(x'))\cdot \frac{\partial f(x')}{\partial x'}$ we get
\begin{eqnarray}
    N_0 &=& \frac{1}{\pi}\int_{-\infty}^{\infty} \frac{\partial f'(x')}{\partial x'} dx', =  \frac{1}{\pi}\int_{-\infty}^{\infty} \frac{dx'}{1+x'^2} \label{Wind1} \\
    &=& \frac{1}{\pi}\tan^{-1}(x')|_{-\infty}^{\infty} = 1.
\end{eqnarray}

\noindent Therefore, at $z=0$ the field winds around the $S_3$ once, this is consistent with the SOP and polarization population plotted in Fig. 1 in the main text.\\ 

\noindent Next, we consider analyzing the winding number on propagation. In Fig. 1 in the main text we observe that the great circle rotates about the $S_1$ axis. In fact, Eq.~\ref{StokesNorm2} and Eq.~\ref{StokesNorm3} are related through 
\begin{equation}
        \vec{S}(x',z) = \hat{R}_{S_1}(z)\vec{S}(x',z=0),
    \label{StokesNormProp1}
\end{equation}

where $ \hat{R}_{S_1}(z) = \begin{bmatrix}
1 & 0 & 0 \\
0 & \cos(\psi(z))&-\sin(\psi(z))\\
0 & \sin(\psi(z)) &\cos(\psi(z)) \\
\end{bmatrix}$ is a z-dependent rotation matrix rotating the Stokes parameters about the $S_1$ axis. We note now that while $S_1$ does not change as the field propagates, $S_2$ decreases while $S_3$ increases. Therefore, if we consider evaluating the winding around the $S_3$ axis, we note that in propagation the vector field $(S_1,S_2)^T$ no longer traces out a circle of unit radius, rather an ellipse with a boundary defined by $g^2 = \cos^2(2f(x')) + \sin^2(2f(x'))\cos^2(\psi(z))$. We note here that while the winding about the $S_3$ axis collapses as $z\to\infty$, since $\lim_{z\to  \infty}S_2(x',z) = 0$, the winding around the $S_2$ axis at this point is exactly the same as that about the $S_3$ axis at $z=0$, i,e, $N^{12}_0 (z=0) = \lim_{z\to\infty} N^{13}_0(z)$. Therefore, the winding number of the great circle is the same at $z=0$ and at $z=\infty$. In between these planes we also anticipate the winding number to remain preserved, since propagation induces a simple rotation of the great circle traced out by the Stokes vector field. We can compute this explicitly however, which alters Eq.~\ref{Wind1} to

\begin{equation}
    N^{12}_0(z) = \frac{\sqrt{b(z)}}{\pi}\int_{-\infty}^{\infty} \frac{1+x'^2 }{x'^4 + 2x'^2 (2b(z)-1) + 1} dx'\label{Wind2}, 
\end{equation}

where $b(z) = \cos(\psi(z))$
Since this integral is symmetric about the $x'=0$ axis we can evaluate the integral in the domain $[0,\infty)$ and multiply the result by a factor of 2.  Further, making the substitution $t=x' - x'^{-1} \implies dt = 1 + x^{-2}$ yields the simplified integral

\begin{eqnarray}
    N^{12}_0(z) &=& 2\frac{\sqrt{b(z)}}{\pi}\int_{-\infty}^{\infty} \frac{dt}{t^2 + 4b}  \\
    &=&  \frac{2\sqrt{b}}{\pi} \left[\frac{1}{2\sqrt{b}}\tan^{-1}\left(\frac{t}{2\sqrt{b}}  \right) \right]_{-\infty}^{\infty} = 1, \label{Wind3}
\end{eqnarray}

thereby confirming the numerical and experimental results shown in Fig.1 in the main results, the winding number is left unchanged on propagation.\\ 

Lastly, in the case where we switch the spatial modes of the polarization components we find that the $S_1$ component flips sign, which introduces a sign change in $\nabla\Phi_{12}$, which results in a sign change of the winding number. 

\section*{Supplementary:Details of experiment and measurements}

To verify the feasibility of the proposed method for generating vector flat-top beams, the experimental setup shown in Fig.~\ref{fig:Setup} was constructed. A linearly polarized beam with a wavelength of 633 nm emitted by a He–Ne laser was expanded and collimated to form an approximate plane wave using a beam-expansion system consisting of a microscope objective (OL) and lens L1, and was then directed onto a spatial light modulator (SLM). The active area of the SLM was equally divided into two halves, on which computer-generated holograms encoding the spatial modes $\alpha LG_{0}$ and $\sqrt{1-\alpha^2} (LG_{-1}+e^{i\theta}LG_{1})$, were displayed to generate two spatially separated co-propagating spatial modes in reflection off the SLM, while preserving the polarization state of the incident light. The two beams subsequently passed through a 4$f$ system composed of lenses L2 and L3 and an aperture placed at their common Fourier plane, where the aperture selected the first-order diffracted beams. A half-wave plate (HWP) was then used to polarization to diagonal polarization. In the vector-beam generation module, the beams entered a Sagnac interferometer comprising a polarizing beam splitter (PBS) and mirrors M2 and M3. The PBS separated the orthogonally polarized $\boldsymbol{\hat{e}_H}$ and $\boldsymbol{\hat{e}_V}$ components, which propagated in opposite directions and were subsequently recombined to synthesize the vector flat-top beam $U(x,y)$. The measurement system consisted of a HWP, a quarter-wave plate (QWP), a linear polarizer (LP), and a CCD camera, while a second 4$f$ system was employed for spatial filtering of the undesired field copies produced within the interferometer. By placing the CCD camera at the image plane and Fourier plane of the second 4$f$ system, the near-field and far-field distributions of the beam were measured, respectively. By adjusting the orientations of the HWP, QWP, and LP, the horizontal, vertical, diagonal, anti-diagonal, right-circularly polarized, and left-circularly polarized intensity components, denoted by $I_H$, $I_V$, $I_D$, $I_A$, $I_R$, and $I_L$, respectively, were recorded. The corresponding Stokes parameters, $S_0, S_1, S_2$ and $S_3$ were then calculated (with further details given in the proceeding section), 
enabling the reconstruction of the beam intensity and spatial polarization-state distributions at different propagation planes.  This setup therefore provides a stable and comprehensive approach for generating, characterizing, and validating vector flat-top beams. 
\begin{figure}[H]
    \centering
    \includegraphics[width=0.7\linewidth]{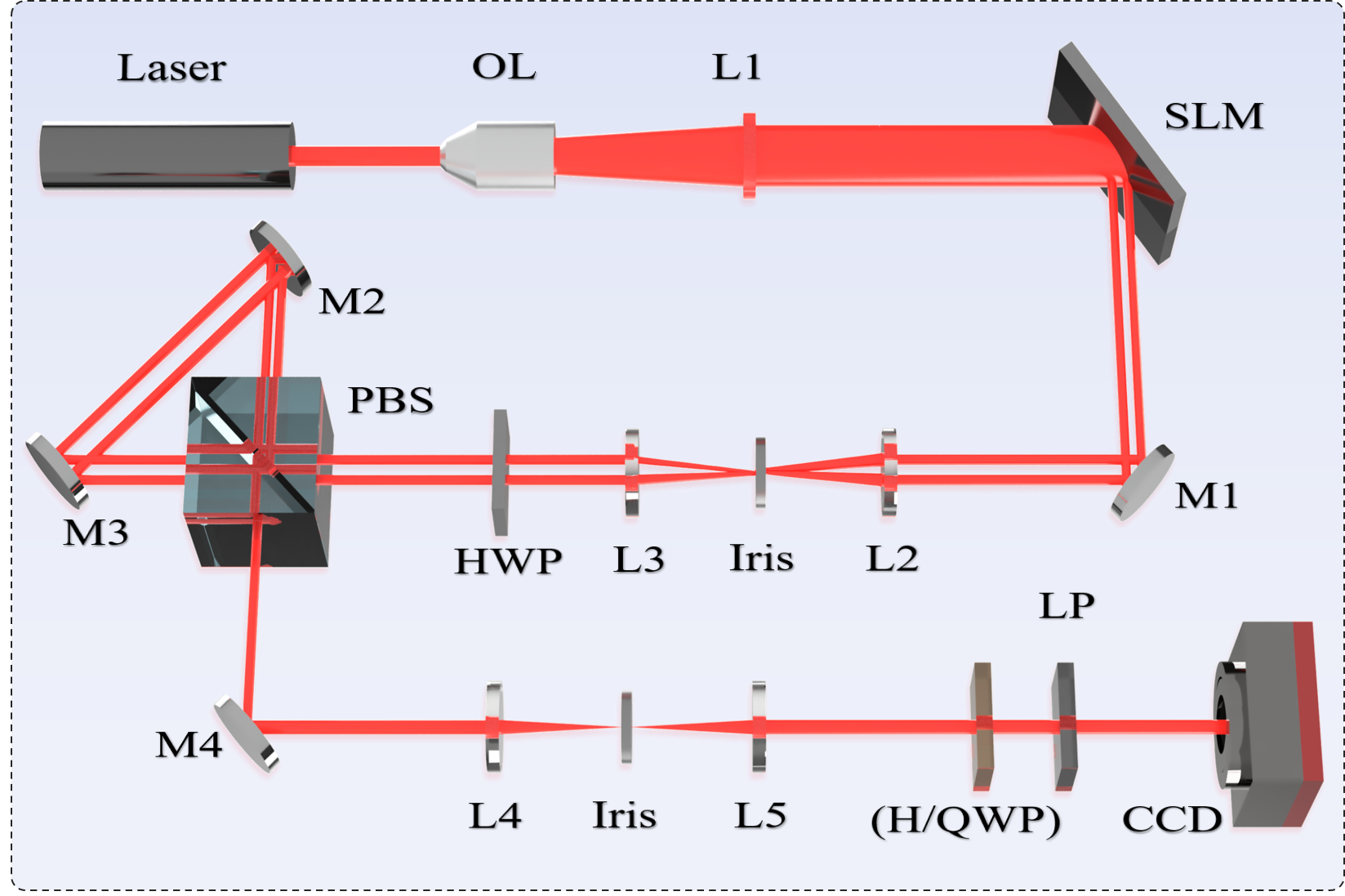}
    \caption{Experimental setup to generate and measure the one-dimensional flat-top vector beam.}
    \label{fig:Setup}
\end{figure}

\section*{Supplementary:Stokes polarimetry and non-separability (vector quality factor) analysis}
To fully characterise a vector beam, the polarisation of the beam must be measured along with the spatial distribution. To measure the polarisation, Stokes polarimetry is performed on the beam to recover the Stokes parameters, which contain all the information about the intensity and polarisation state of the beam. Stokes polarimetry involves measuring the intensity of the 6 polarisation projections: Horizontal,$I_H$; Vertical,$I_V$; Diagonal, $I_D$; Anti-diagonal, $I_A$; Right-circular, $I_R$; and Left-circular polarisations,$I_L$. 
The Stokes parameters are calculated from the intensity measurements as 
\begin{equation}
    \begin{split}
        S_0 &= I_H + I_V\\
        S_1 &= I_H-I_V\\
        S_2 &= I_D - I_A\\
        S_3 &= I_R - I_L
    \end{split}
\end{equation}
Fig.~\ref {fig: Stokes in prop} shows experimental measurements of the Stokes parameters. The Stokes parameters were measured at different propagation planes $z = nz_R$, where $n\in[0, \tfrac{1}{4}, \tfrac{1}{2}, \tfrac{3}{4}, 1]$ and finally at $z>>z_R$. While $S_0$ and $S_1$ remain the same, with rescaled sizes, in propagation, $S_2$ decreases in intensity while $S_3$ increases. This corroborates the result shown in Fig. 1 (d) of the main text; the polarization loop rotates about $S_1$ as the magnitude of $S_2$ decreases and $S_3$ increases.    
Additionally, we calculate the degree of vectorness (VQF) of the beam from the Stokes parameters as follows 
\begin{equation}
    VQF = \sqrt{1- \left(\dfrac{\sigma_1^2 + \sigma_2^2 + \sigma_3^2}{\sigma_0^2}\right)},
\end{equation}
where, $\sigma_i = \int S_idA$.
To calculate the VQF of the 1D flattop field described in \ref{Field eq init}, we first describe the polarisation states in terms of their Jones vectors as 
\begin{equation}
\begin{alignedat}{4}
\boldsymbol{\hat{e}_H} &= \begin{pmatrix} 1 \\ 0 \end{pmatrix};
&\qquad
\boldsymbol{\hat{e}_V} &= \begin{pmatrix} 0 \\ 1 \end{pmatrix} ;\\[0.8em]
\boldsymbol{\hat{e}_D} &= \dfrac{1}{\sqrt{2}}\begin{pmatrix} 1 \\ 1 \end{pmatrix};
&\qquad
\boldsymbol{\hat{e}_A} &=\dfrac{1}{\sqrt{2}} \begin{pmatrix} 1 \\ -1 \end{pmatrix}; \\[0.8em]
\boldsymbol{\hat{e}_R} &= \dfrac{1}{\sqrt{2}}\begin{pmatrix} 1 \\ i \end{pmatrix};\text{and }
&\qquad
\boldsymbol{\hat{e}_L} &= \dfrac{1}{\sqrt{2}}\begin{pmatrix} 1 \\ -i \end{pmatrix}.
\end{alignedat}
\end{equation}
where $\boldsymbol{\hat{e}_H}$ and $\boldsymbol{\hat{e}_V}$ are the basis vectors. To simplify the calculation, we will integrate the intensity over the whole field such that the spatial distributions integrate to 1 and the intensities reduce to 
\begin{equation}
\begin{alignedat}{4}
I_H &= |\alpha|^2;
&\qquad
I_V &= |\sqrt{1-\alpha^2}|^2 ;\\[0.8em]
I_D &= |\frac{\alpha}{\sqrt{2}}+ \frac{\sqrt{1-\alpha^2}}{\sqrt{2}}|^2  ;
&\qquad
I_A &=|\frac{\alpha}{\sqrt{2}} - \frac{\sqrt{1-\alpha^2}}{\sqrt{2}}|^2; \\[0.8em]
I_R &= |\frac{\alpha}{\sqrt{2}} +i \frac{\sqrt{1-\alpha^2}}{\sqrt{2}}|^2;\text{and }
&\qquad
I_L &= |\frac{\alpha}{\sqrt{2}} -i \frac{\sqrt{1-\alpha^2}}{\sqrt{2}}|^2.
\end{alignedat}
\end{equation}
For the case of $\alpha  = \frac{1}{\sqrt{3}}$, the Stokes parameters can be calculated to be $(S_0, S_1, S_2, S_3) = (1 , -\frac{1}{3}, 0, 0)$
and $\textbf{VQF} = \sqrt{1- (\frac{1}{9})} = 0.94$.

\begin{figure}[H]
    \centering
    \includegraphics[width=\linewidth]{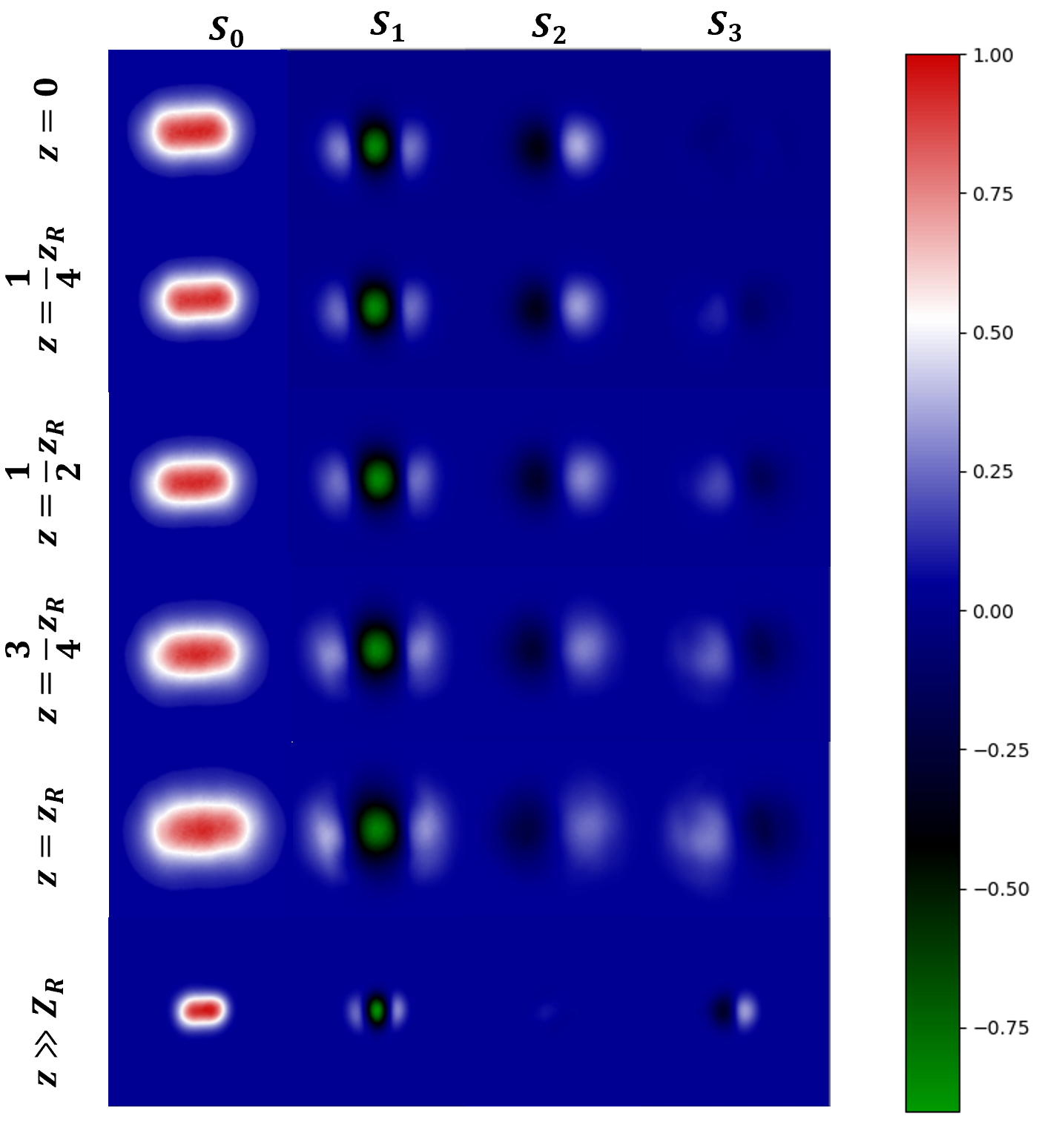}
    \caption{Stokes parameters $S_0, S_1, S_2$ and $S_3$ measured at different propagation planes}
    \label{fig: Stokes in prop}
\end{figure}
\end{widetext}

\end{document}